\documentclass[11pt,twoside]{article}
\usepackage{./asp2014}

\aspSuppressVolSlug
\resetcounters

\begin{document}

\title{Gas flows in galactic centre environments: cloud evolution and star formation in the Central Molecular Zone}
\author{Maya~A.~Petkova and J.~M.~Diederik~Kruijssen
\affil{Astronomisches Rechen-Institut, Zentrum f\"ur Astronomie der Universit\" at Heidelberg,  M\"onchhofstra\ss e 12-14, 69120 Heidelberg, Germany;\\
\email{maya.petkova@uni-heidelberg.de}}}

% This section is for ADS Processing.  There must be one line per author.
\paperauthor{Maya~A.~Petkova}{maya.petkova@uni-heidelberg.de}{0000-0002-6362-8159}{Zentrum f\"ur Astronomie der Universit\" at Heidelberg}{Astronomisches Rechen-Institut}{Heidelberg}{Baden-Wuerttemberg}{69120}{Germany}
\paperauthor{J.~M.~Diederik~Kruijssen}{kruijssen@uni-heidelberg.de}{0000-0002-8804-0212}{Zentrum f\"ur Astronomie der Universit\" at Heidelberg}{Astronomisches Rechen-Institut}{Heidelberg}{Baden-Wuerttemberg}{69120}{Germany}

\begin{abstract}
The Central Molecular Zone (CMZ; the central ~500 pc of the Galaxy) is the most extreme star-forming environment in the Milky Way in terms of gas pressures, densities, temperatures, and dynamics. It acts as a critical test bed for developing star formation theories applicable to the (high-redshift-like) conditions under which most stars in the Universe formed. We present a set of numerical simulations of molecular clouds orbiting on the 100-pc stream that dominates the molecular gas reservoir of the CMZ, with the goal of characterising their morphological and kinematic evolution in response to the external gravitational potential and their eccentric orbital motion. These simulations capture the evolution of single clouds in a strong and plausibly dominant background potential. We find that the evolution of the clouds is closely coupled to the orbital dynamics and their arrival on the 100-pc stream marks a transformative event in their lifecycle. The clouds' sizes, aspect ratios, position angles, filamentary structure, column densities, velocity dispersions, line-of-sight velocity gradients, angular momenta, and overall kinematic complexity are controlled by the background potential and their passage through the orbit's pericentre. We compare these predictions of our simulations to observations of clouds on the Galactic Centre `dust ridge' and find that the inclusion of galactic dynamics naturally reproduces a surprisingly wide variety of key observed morphological and kinematic features. We argue that the accretion of gas clouds onto the central regions of galaxies, where the rotation curve turns over and the tidal field becomes fully compressive, is likely to lead to their collapse and associated star formation. This can generate an evolutionary progression of cloud collapse with a common zero point. Together, these processes may naturally give rise to the synchronised starbursts observed in numerous (extra)galactic nuclei.
\end{abstract}

%\section{Introduction}
%Give some background to why you're doing this.

\section{Background}
The `dust ridge' is a stream of clouds in the CMZ occupying an eccentric orbit between 60 and 120 pc (Molinari et al. 2011; Longmore et al. 2013b; Kruijssen et al. 2015; Henshaw et al. 2016). The clouds there are subjected to an extreme dynamical environment of a fully compressive tidal field (Kruijssen et al. 2015; Kruijssen et al. 2019) and strong shear (Krumholz \& Kruijssen 2015; Krumholz et al. 2017; Jeffreson et al. 2018), generated by the gravitational potential. This dynamical environment may explain why the CMZ clouds violate galactic and extragalactic star formation relations, as they are underproducing stars for their densities (Longmore et al. 2013; Kruijssen et al. 2014; Barnes et al. 2017; Kauffmann et al. 2017). What is also unique and interesting about these clouds is that they have high pressures, densities and temperatures, which are extreme compared to the typical values found in the Galactic disc, but are similar to those in high-redshift galaxies (Kruijssen \& Longmore 2013; Ginsburg et al. 2016). This means that studying star formation in the CMZ gives us an opportunity to better understand star formation in the early Universe. 

\section{Smoothed particle hydrodynamics (SPH) simulations of CMZ clouds}
To investigate the impact of galactic dynamics on the CMZ clouds, we present five SPH simulations of CMZ clouds: a fiducial one, high and low density ones, and high and low velocity dispersion ones, originally presented by Kruijssen et al. (2019) and Dale et al. (2019). The clouds follow the open eccentric orbit described by Kruijssen et al. (2015) and Henshaw et al. (2016). Along their orbit, the clouds experience a static gravitational potential, which is empirically derived from the stellar distribution dominating the mass in the region (Launhardt et al. (2002); Kruijssen et al. 2015). Each cloud has been evolved individually, in a simulation accounting for the self-gravity of the gas but disregarding any effects of stellar feedback.

The simulations provide a good match to the observed `dust ridge' in terms of their morphology (see Figure 1). The simulated clouds naturally reproduce observables such as the aspect ratio, column density and velocity dispersion, which directly results from their orbital evolution within the gravitational potential of the Galactic Centre environment. The combined effects of shear and fully compressive tides (both set by the potential) cause the clouds to become vertically compressed, while they maintain larger size in their orbital plane. This creates the elongated cloud shapes that we observe. The pericentre passage is a point of orbit compression, and as such it boosts the clouds' ongoing gravitational collapse and star formation (Dale et al. 2019).

\articlefigure{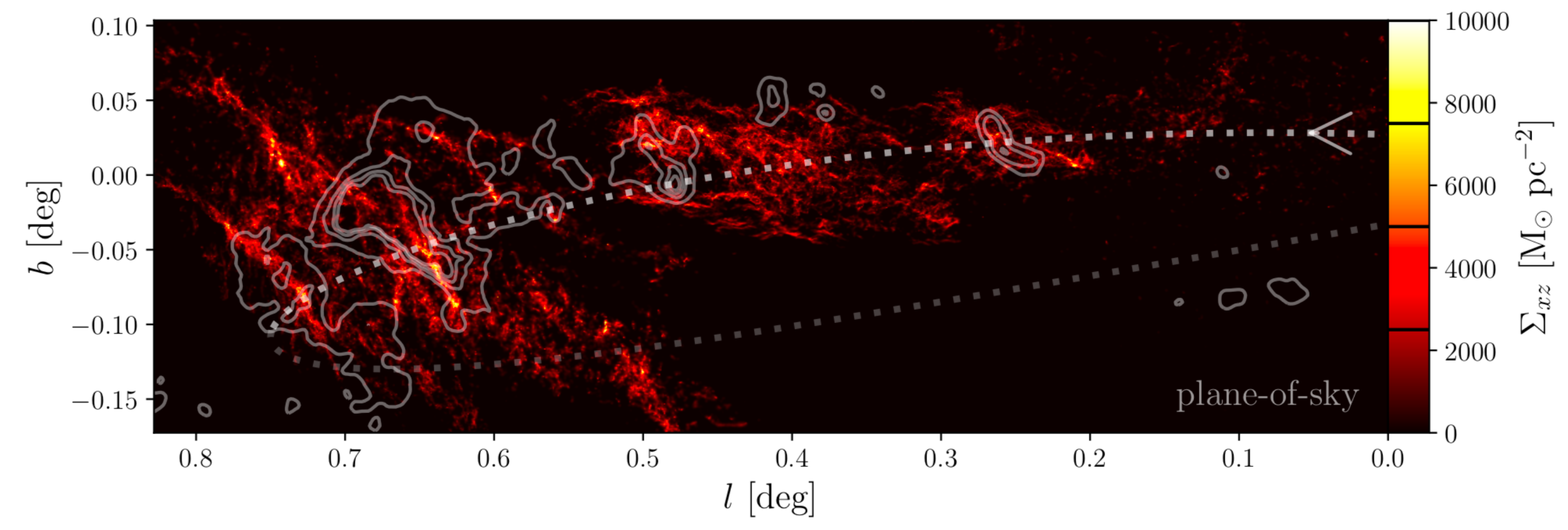}{fig1}{Column density map of the simulated `dust ridge' clouds of the CMZ (adapted from Figure 2 in Kruijssen et al. 2019). Snapshots of the different simulated clouds are shown together at chosen positions along their orbit and viewed in the plane of the sky. The dotted line shows the orbital solution of Kruijssen et al. (2015), with an arrow indicating the direction of motion. Observed column density maps from HiGAL (see Battersby et al. in prep.) are shown as contours at \{2.5, 5, 7.5, $10\}\times 10^3$ M$_\odot$ pc$^{-2}$.}

Initially, the clouds are given turbulent velocities that result in a small amount of counter-rotation. As they start moving along their orbit, the turbulence is gradually dissipated (as it is not explicitly driven in the simulations, other than by any shear-generated motion of the clouds), which results in a decrease of the line-of-sight velocity dispersion. After the pericentre passage, gravitational collapse sets in and the line-of-sight velocity dispersion increases again. At its minimum, however, the velocity dispersion does not drop to zero, but instead remains at a floor consistent with the effects of shear. This suggests that the strong shear in the CMZ can indeed drive at least some of the turbulence in the observed dust ridge clouds. 

In addition to the above results, we have analysed the detailed morphology of the simulation corresponding to G0.253+0.016, also known as the Brick. A visual comparison between the filamentary structure of the simulated and observed cloud is shown in Figure 2. In order to expand on the work of Kruijssen et al. (2019) and quantify the morphological complexity, we have also created a synthetic ALMA observation and we have calculated its fractal dimension (Petkova et al. in prep.). This is achieved by post-processing the simulation with the line radiative transfer code POLARIS (Reissl et al. 2016) and further processing the resulting image with CASA (McMullin et al. 2007). The fractal dimension is obtained by selecting connected regions of pixels with values over a chosen threshold, and fitting the relation between their perimeter and area. The resulting slope implies a fractal dimension of 1.4, which agrees with the range of 1.4--1.7 measured by Rathborne et al.(2015).

\articlefigure{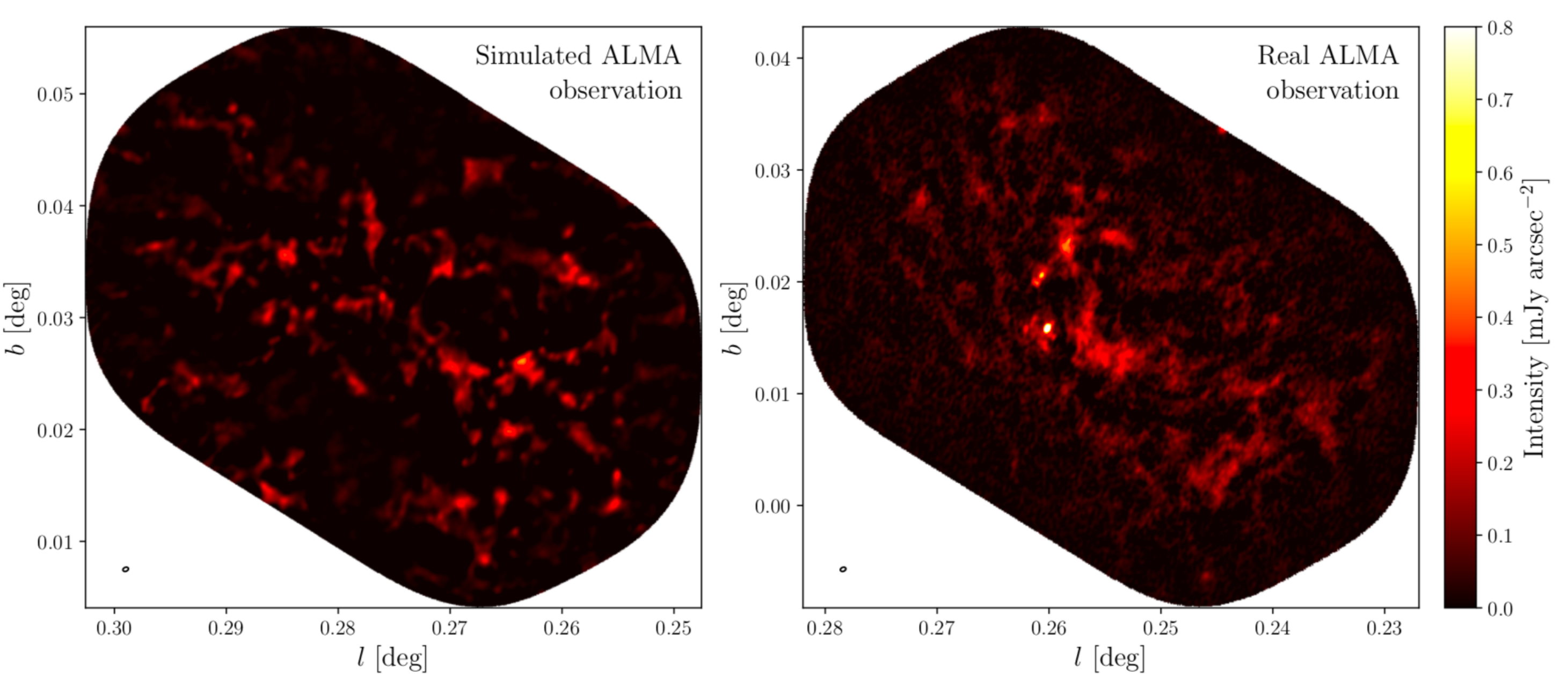}{fig2}{Comparison between a simulated \textit{(left)} and real \textit{(right)} ALMA observation of the Brick (adapted from Figure 6 in Kruijssen et al. 2019). Both images use the same colour scale and the ellipses in the bottom left corner indicate the beam size. The observational data is taken from Rathborne et al.(2015).}

Finally, we also study the effects of ionising stellar feedback on the simulation snapshot best reproducing the Brick. This is achieved with the use of a radiation-hydrodynamics scheme presented in Petkova et al. 2018 and Petkova et al. in prep. We use two ionising sources --- one embedded in a clump and one in low-density environment. Even though these ionising feedback models are simplified by not including gravitational forces, we find that both setups can influence the cloud dynamics on short timescales (< 0.1 Myr) by ionising the low- and medium-density gas in the vicinity of the source. In the case of the embedded source the models require very high ionising luminosities (corresponding to a star of > 100 M$_{\odot}$) in order for the ionised region to expand beyond the boundary of the clump. This can explain why we have not observed an H II region in the real Brick.

\section{Conclusions}
We have presented hydrodynamical simulations of molecular clouds in the CMZ to study the effects of the external potential on the cloud evolution. We find that CMZ clouds are shaped by shear, tidal and geometric deformation, and their passage through orbital pericentre. Furthermore, we find evidence that shear is an important turbulence driver in the CMZ.
The large velocity gradients observed in the `dust ridge' clouds can be explained by gravitational collapse, and the observed aspect ratio is a result of the compressive tidal field and the shear. Additionally, we find that the simulated morphology of the Brick is consistent with observations, and that strong ionising stellar feedback can alter the structure of the cloud on a short time scale. The absence of such effects in the observed cloud implies that the Brick has not undergone any massive star formation prior to the past 0.1 Myr.

\acknowledgements We thank Jim Dale, Steve Longmore, Daniel Walker, Jonathan Henshaw, Sarah Jeffreson, Adam Ginsburg and Simon Glover for their contributions. We are supported by the European Research Council (ERC; ERC-StG-714907). JMDK is supported by the German Research Foundation (DFG; KR4801/1-1). 

%\bibliography{editor}  % For BibTex

% For non-BibTex:

\end{document}